\documentclass[preprint]{imsart}

\usepackage{amsthm,amsmath,amssymb,mathrsfs,bm}
\RequirePackage[colorlinks,citecolor=blue,urlcolor=blue]{hyperref}
\numberwithin{equation}{section}
\newtheorem{theorem}{Theorem}[section]
\newtheorem{lemma}[theorem]{Lemma}
\newtheorem{corollary}[theorem]{Corollary}
\newtheorem{proposition}[theorem]{Proposition}

\newtheorem{remark}[theorem]{Remark}

\def\pp{{\bf P}}
\def\ppn{{\bf P}_{\hspace{-2pt}n}}

\def\dd{{\rm d}}
\def\qq{{\bf Q}}

\def\NF{\mathbb{F}}
\def\NG{\mathbb{G}}
\def\NH{\mathbb{H}}

\def\NR{\mathbb{R}}

\begin{document}

\begin{frontmatter}

\title{On the collapse of trial solutions\\ for a
damped-driven\\ non-linear Schr\"odinger equation}
\runtitle{Collapse of Trial Solution}

\author{\fnms{Sigurd} \snm{Assing}\corref{}\ead[label=e1]{s.assing@warwick.ac.uk}}
\address{Department of Statistics\\
University of Warwick\\
\printead{e1}}
\affiliation{University of Warwick}

\author{\fnms{Astrid} \snm{Hilbert}\ead[label=e2]{astrid.hilbert@lnu.se}}
\address{Department of Mathematics\\
Linn\'euniversitetet Vaxjo\\
\printead{e2}}
\affiliation{Linn\'euniversitetet Vaxjo}

\runauthor{Assing and Hilbert}

\begin{abstract}
We consider the 
focusing 2D non-linear Schr\"odinger equation,
perturbed by a damping term, and driven by multiplicative noise.
We show that a physically motivated trial solution
does not collapse for any admissible initial condition
although the exponent of the non-linearity is critical.
Our method is based on the construction of a global solution
to a singular stochastic Hamiltonian system used to connect 
trial solution and Schr\"odinger equation.
\end{abstract}



\end{frontmatter}

\section{Motivation}
Consider the formal equation,
\begin{equation}\label{NLS}
{\bf i}\,\partial_t\psi
+\Delta_u\psi+|\psi|^2\psi-\Lambda\,\partial_t(|\psi|^2)\psi
+\sigma(u,t)\psi\,=\,0,
\end{equation}
with Cauchy data at $t=0$, where $[\sigma(u,t),\,(u,t)\in\NR^2\times(0,\infty)]$ is 
radially symmetric centred Gaussian noise with covariance
\[
\langle\sigma(u,t)\sigma(u',t')\rangle
\,=\,
\frac{D_r}{|u|}\,\delta_0(|u|-|u'|)\delta_0(t-t').
\]
This equation was derived in \cite{main} as the isotropic
continuum approximation of a model for two-dimensional
damped-driven exciton-phonon systems.

Note that (\ref{NLS}), as derived in Section II of \cite{main},
is actually driven by coloured multiplicative noise.
But, in Section III of \cite{main}, the authors say they 
would rather approximate the driving noise by space-time white noise
which they had justified in \cite{main(3)}.
Finally, in order to allow radially symmetric (i.e.\ isotropic) solutions,
they simplified space-time white noise to radially symmetric Gaussian noise
as used in the formulation of (\ref{NLS})---the reader is referred to \cite{main(3)}
for the definition of a parameter $D_{white}$ which can be used to choose
a physically meaningful value for $D_r$. The definition, in physical terms,
of the positive damping parameter $\Lambda$ can be found in \cite{main}, too.

In the case of $\Lambda=D_r=0$, equation (\ref{NLS}) is identical to the
classical focusing (power) non-linear Schr\"odinger equation,
and the power {\it two} in the non-linearity $|\psi|^2\psi$
is known to be the smallest power-like non-linearity for which blow-up occurs
in space dimension $d=2$.

For example, the wave function,
\begin{equation}\label{special blow up}
\psi(u,t)\,=\,
|t-1|^{-\frac{d}{2}}Q(\frac{u}{t-1})\,e^{{\bf i}\frac{|u|^2}{4(t-1)}-\,{\bf i}/(t-1)},
\end{equation}
where $Q$ is the unique positive radially symmetric solution of
\[
\Delta Q-Q+Q^{1+\frac{4}{d}}\,=\,0\;\;\mbox{in $\NR^d$},
\quad\mbox{with $Q(u)\to 0$ as $|u|\to\infty$},
\]
can be viewed as a solution of
\[
{\bf i}\,\partial_t\psi+\Delta_u\psi+|\psi|^{\frac{4}{d}}\psi\,=\,0,
\quad\mbox{in $\NR^d\times(0,\infty)$},
\]
with Cauchy data,
\[
\psi(u,0)\,=\,Q(u)\,e^{-{\bf i}(\frac{|u|}{4}^{\!2}-1)},
\quad\mbox{at $t=0$},
\]
which blows up at time $t=1$. Note that $Q$ is also called {\it ground state}.

Now observe that
\[
\int_{\NR^d}|\psi(u,t)|^2\,\dd u\,=\int_{\NR^d}Q(u)^2\,\dd u,\quad
\mbox{for all $t\in[0,1)$},
\]
in the above example.
Hence, all $L^2$\,-\,mass is accumulated into blow-up, and, by the shape of $Q$,
this $L^2$\,-\,mass is concentrated at $u=0$ at time $t=1$.

The above described phenomenon, also called the $L^2$\,-\,concentration phenomenon,
is well-known for $L^2$\,-\,critical Schr\"odinger equations---the reader
is referred to \cite{merle1},\cite{merle2},\cite{nawa} for general results.

Now, in \cite{main},
the authors have asked if this phenomenon was possible for solutions
of their model for damped-driven exciton-phonon coupled systems,
as the balanced energy input could prevent solutions from blow-up.

Of course, (\ref{NLS}) is hard to solve, and being a formal equation only,
its rigorous meaning would need further discussion in the first place.
This difficulty was by-passed in \cite{main}. Instead, the authors
introduced the following family of wave functions,\hspace{-2pt}\footnote{
See Remark \ref{parameters} for the definition of $c_f^{m,n,p}$.} 
\begin{equation}\label{trial wave}
\psi(u,t)\,\stackrel{\mbox{\tiny def}}{=}\,
\frac{\|\psi(\cdot,0)\|_{L^2}}{\sqrt{c_f^{1,2,0}}}\times
\frac{1}{x(t)}\,
f(\frac{|u|}{x(t)})\,e^{{\bf i}\frac{\dot{x}(t)|u|^2}{4x(t)}},
\end{equation}
parametrised by $\|\psi(\cdot,0)\|_{L^2}$, 
a smooth function $f:\NR\to(0,\infty)$ which is rapidly decreasing,
and an unknown stochastic process $x=[x(t),\,t\ge 0]$
which plays the role of the width of the corresponding non-linear wave.

Note that $f$, in contrast to $Q$ used in (\ref{special blow up}),
does not have to satisfy any equation.
Nevertheless, similar to (\ref{special blow up}), 
wave functions of this type would blow up,
if $x$ starting from a positive value hits zero in finite time,
the initial $L^2$\,-\,mass being preserved in the process.
Due to the nature of a blow-up with vanishing width of the wave function,
we also call such a behaviour {\it collapse}.

So, the question asked can be scaled down to the following problems:
\begin{itemize}
\item[a)]
choose $x$ in a way such that the trial solution (\ref{trial wave}) 
has something to do with the primary equation (\ref{NLS});
\item[b)]
study whether $x$ chosen this way reaches zero in finite time or not.
\end{itemize}

To shed some light on a), restricting ourselves to isotropic solutions,
we can rewrite (\ref{NLS}) as
\[
{\bf i}\,\partial_t\psi
+\partial_r^2\psi+\frac{1}{r}\partial_r\psi+
|\psi|^2\psi-\Lambda\,\partial_t(|\psi|^2)\psi
+\sigma(r,t)\psi\,=\,0,
\]
reducing the problem to one space dimension with radial coordinate $r=|u|$.
Since $\sigma$ has negative H\"older-regularity,
non-trivial solutions of this equation are not smooth in $r$,
but the trial solution (\ref{trial wave}) is.
Therefore, (\ref{trial wave}) can at most be 
an approximate solution having some features of a true solution.

The features chosen in \cite{main} involve the virial coefficient
\[
{\mathfrak v}(t)\,=\,
\int_{\NR^2}|u|^2|\psi(u,t)|^2\,\dd u\,=\,
2\pi\int_0^\infty r^3\,|\psi(r,t)|^2\,\dd r
\]
which,
when $\psi$ is assumed to solve (\ref{NLS}),
would formally satisfy
\begin{equation}\label{virial}
\ddot{\mathfrak v}\,=\,
16\mathscr{H}+8\pi\int_0^\infty r^2\,|\psi|^2\,
\partial_r[-\Lambda\partial_t(|\psi|^2)+\sigma]\,\dd r
\end{equation}
where
\[
\mathscr{H}(t)\,=\,
\int_{\NR^2}
\left[\rule{0pt}{12pt}\right.
\frac{1}{2}\,|\nabla_u\psi(u,t)|^2
-
\frac{1}{4}\,|\psi(u,t)|^4
\left.\rule{0pt}{12pt}\right]
\dd u
\]
is the Hamiltonian 
of the focusing cubic non-linear Schr\"odinger equation.

Plugging $\psi$ given by (\ref{trial wave}) into (\ref{virial}),
and then performing the integration against $r$, transforms (\ref{virial})
into an equation for $x$, only.
The authors of \cite{main} now argue 
that this equation can be studied by the simpler equation,
\begin{equation}\label{single equ}
\ddot{x}\,=\,\frac{\delta}{x^3}
+[\sqrt{2D}\,\frac{\dot{W}}{x^2}-\frac{\gamma\dot{x}}{x^4}\,],
\end{equation}
where $W=[W(t),\,t\ge 0]$ stands for a standard one-dimensional Wiener process.
\begin{remark}\rm\label{parameters}
As stated in \cite{main},
\[
\delta\,=\,\frac{4}{c_f^{3,2,0}}\,
\left(\rule{0pt}{12pt}\right.
c_f^{1,0,2}
-\frac{\|\psi(\cdot,0)\|_{L^2}^2\,c_f^{1,4,0}}{2c_f^{1,2,0}}
\left.\rule{0pt}{12pt}\right),
\]
and
\[
\gamma\,=\,
{8\Lambda\|\psi(\cdot,0)\|_{L^2}^2\,c_f^{3,2,2}}/(c_f^{1,2,0}c_f^{3,2,0})
,\quad
D\,=\,{32\pi^2 D_r\,c_f^{3,2,2}}/{(c_f^{3,2,0})^2},
\]
using
\[
c_f^{m,n,p}
\,\stackrel{\mbox{\tiny def}}{=}\,
2\pi\int_0^\infty r^m[f(r)]^n[f'(r)]^p\,\dd r.
\]
\end{remark}

The methodology for solving the first problem, a),
as developed in \cite{main}), can therefore be described as follows:
for fixed $\|\psi(\cdot,0)\|_{L^2}$ and $f$,
fit the trial solution (\ref{trial wave}) to (\ref{virial})
to obtain equation (\ref{single equ}) for the unknown $x$.
This way, equation (\ref{NLS}) is more or less replaced by (\ref{virial})
subject to a structure condition---the specific form of $\psi$ 
given by (\ref{trial wave})---which may put such a $\psi$
in close vicinity of a true solution to (\ref{NLS}),
in particular for well-chosen $f$.

However, no rigorous analysis has been done to support such a quality 
of the trial solution (\ref{trial wave}). 
First, one would have to make rigorous sense of all formal calculations
used to motivate both equations (\ref{virial}) \& (\ref{single equ}), 
and, second, one would have to study
how close the trial solution (\ref{trial wave}) and 
true solutions to (\ref{NLS}) really are, and in what sense.

These open but very interesting problems are beyond the present paper
and left for future research. We should nevertheless mention that
numerical experiments reported in physics journals confirm a close match
of trial solutions and true solutions on short time intervals.
So, studying the long-time behaviour of solutions to (\ref{single equ}),
and in particular answering the second problem, b), seems to be a natural
next step in the analysis of the original problem. For example, once
the answer to b) is known, blow-up of true solutions could be decided 
by merely comparing functionals of trial solutions and true solutions.

In this paper, we therefore study the physically motivated
equation (\ref{single equ}) and solve the second problem, b),
for all parameters of interest.
Apart from the above application,
these results are interesting in themselves since equation (\ref{single equ}) 
describes the dynamics of a perturbed stochastically driven Hamiltonian system
with a singular potential (cf.\ \cite{mattingly,funaki} and Remark \ref{potentials} below).
\begin{remark}\rm
\rule{1cm}{0pt}
\begin{itemize}\item[(i)]
Although the trial solution (\ref{trial wave}) 
has infinite degrees of freedom,
studying it's blow-up through (\ref{single equ}) means 
that only the interplay of five one-dimensional parameters has to be considered:
$x(0),\dot{x}(0),\delta,\gamma,D$.
Here, assuming $\Lambda,D_r>0$, one also has
$\gamma,D>0$ by Remark \ref{parameters}, 
and since the trial solution has a physical meaning,
$x(0)>0$ should hold, too.
\item[(ii)]
Note that the parameter $\delta$ can have both signs depending on the relationship
between $\|\psi(\cdot,0)\|_{L^2}^2$ and integrals of $f$.
Of course, compared with negative $\delta$,
the width $x$ of the wave function is less likely to reach zero in finite time 
when $\delta$ is positive, and hence
one might expect a phase transition in the behaviour of solutions to (\ref{single equ})
depending on $\delta$.
However, we are going to prove that, for any choice of $\delta$,
the width $x$ never collapses to zero in finite time.
\end{itemize}
\end{remark}
\section{Results}\label{results}
Motivated by (\ref{single equ}),
we study the degenerated It\^{o} diffusion equation,
\begin{equation}\label{strato diff}
\begin{array}{rcl}
\dd x&=&y\,\dd t\,,\\
\dd y&=&\displaystyle
\frac{a}{x^\alpha}\,\dd t
+[\,\frac{\sqrt{2T\gamma}}{x^\beta}\,\dd W-\frac{\gamma}{x^{2\beta}}\,y\,\dd t\,],
\end{array}
\end{equation}
where $W$ denotes a one-dimensional Wiener process.

Our goal is to construct a global solution $[x(t),y(t),\,t\ge 0]$,
for any initial condition 
$(x(0),y(0))\in\NH\stackrel{\mbox{\tiny def}}{=}(0,\infty)\times\NR$,
if both $\gamma,T>0$, but $a\in\NR$ is a real parameter with no restrictions on its sign.

We also assume $\beta>1/2$ for reasons explained in Section \ref{why} below.

Furthermore, let $\alpha>1$, though this  will be relaxed for $a\ge 0$, later.

First, note that
the infinitesimal operator associated with this equation can formally be written as
\begin{equation}\label{infi op}
{\cal L}
\,=\,
\frac{\partial H}{\partial y}\partial_x
-\frac{\partial H}{\partial x}\partial_y
+[\,T\frac{\gamma}{x^{2\beta}}\partial_y^2
-\frac{\gamma}{x^{2\beta}}y\,\partial_y\,],
\end{equation}
using the Hamiltonian,
\[
H(x,y)\,=\,U(x)+\frac{y^2}{2},
\quad\mbox{with potential}\quad
U(x)\,=\,\frac{a}{\alpha-1}\,x^{1-\alpha},
\]
so that (\ref{strato diff}) can be interpreted
as a damped-driven Hamiltonian system,
being forced by noise on an $x$-depending scale
which exactly balances the $x$-depending dissipation
placing the stochastic system at temperature $T$.

The difficulty we are facing is twofold: 
first, our potential $U$ has a singularity at zero, and, second, 
{\it intensity of noise} = $\sqrt{2T\times\mbox{\it friction}}$\, = $\sqrt{2T\times\gamma/x^{2\beta}}$ 
not only depends on $x$, but has a singularity at zero, too. 

Nevertheless, because noise and dissipation are balanced, 
\[
\varrho_\star(x,y)\,=\,e^{-H(x,y)/T}
\quad\mbox{satisfies}\quad
{\cal L}^\star\varrho_\star=0\;\mbox{on $\NH$},
\]
where ${\cal L}^\star$ stands for the (formal) adjoint of ${\cal L}$.
However, in our case,
\begin{equation}\label{divergence}
\int_0^\infty\hspace{-5pt}\int_\NR e^{-H(x,y)/T}\,\dd y\dd x\,=\,+\infty,
\end{equation}
which means that this density cannot be normalised to become the density
of the system's canonical invariant probability measure. 
\begin{remark}\rm\label{potentials}
\rule{1cm}{0pt}
\begin{itemize}\item[(i)]
There is recent work in \cite{mattingly}
on stochastic dynamical systems associated with operators of type (\ref{infi op})
where singular potentials of the form,
\[
U(x)\,=\,
a_1 x^{\alpha_1}+a_2 x^{-\alpha_2},
\]
are considered
assuming $a_1,a_2>0,\,\alpha_1>2$, 
but where intensity of noise and friction are constant, i.e.\ $\beta=0$.
Here, the integral (\ref{divergence}) converges, and 
the canonical invariant measure occurs to be the system's unique invariant measure.
More importantly, the authors describe a general method of how to construct
a Lyapunov function which gives 
control over the system's trajectories near the boundary of $\NH$.
Note that their parameter $a_2$---the analogue of our parameter $a$---is 
assumed to be positive, which makes their potential repulsive at zero.
If our potential is repulsive at zero, i.e.\ $a>0$, we are going to show that
$\lim_{t\to\infty}x(t)=+\infty$, 
and hence a Lyapunov function cannot exist.
\item[(ii)]
The driven Rayleigh-Plesset equation considered in \cite{funaki}
has features being more similar to our equation (\ref{strato diff}).
Only looking at the degenerate-diffusion-case
(cf.\ Section 5 in \cite{funaki}),
their potential has no $a_1$-term either,
and the leading singularity is of the form $b x^{-3k}$, for some $k\ge 1$, 
but again $b>0$, so that the singular
potential is repulsive at zero, as in \cite{mattingly}. Intensity of noise
and friction, though, depend on both $x$ and $y$, but in an unbalanced way.
The special form of the unbalanced noise and dissipation terms
together with the repulsive potential make it possible to find 
a Lyapunov function, and thus an invariant measure exists.
\item[(iii)]
It should be mentioned that the methods used in both papers \cite{mattingly}
and \cite{funaki} will not work when the potential is not repulsive at zero.
\end{itemize}
\end{remark}

All in all, 
results on stochastic dynamical systems as singular as ours
were only obtained when the system 
admitted a structure inducing contraction properties 
like existence of a Lyapunov function---see \cite{mattingly} for further references.
As our system (\ref{strato diff}) does not have such a structure,
a different technique is needed, even for showing existence of global solutions.

Let us briefly discuss the deterministic case, i.e.\ $\gamma=0$. Here,
${\cal L}H=0$ on the upper half-plane $\NH$,
and thus, if $a>0$, 
then $f_L(x,y)=H(x,y)+x^2/2$ is a non-negative function on $\NH$,
satisfying
\[
\lim_{(x,y)\to\partial\NH}\;f_L(x,y)\,=\,+\infty
\quad\mbox{and}\quad
{\cal L}f_L\,\le\,const\,(1+f_L).
\]
Since the coefficients of equation (\ref{strato diff}) are locally Lipschitz on $\NH$,
by standard arguments, for any $(x(0),y(0))\in\NH$,
there is a global solution $[x(t),y(t),\,t\ge 0]$ 
whose first component never reaches zero in finite time.

However, if $a<0$, then there is no function $f_L$ as above.
Worse, the $x$-component of any solution in $\NH$
eventually collapses to zero in finite time, 
and hence there is no global solution.

So, the question is:
can adding noise but also dissipation,
both balancing each other as in equation (\ref{strato diff}), 
prevent collapse of solutions in $\NH$, even for $a<0$? The answer is YES 
for certain choices of the exponents $\alpha,\beta$ depending on
whether $a>0,\,a=0$, or $a<0$.

The proof of our existence result is based on the following lemma
which is the key-result of this paper.
\begin{lemma}\label{technical}
Assume $\gamma,T>0$, and
\begin{equation}\label{?}
\left.
\begin{array}{crcl}
\bullet&\beta>1/2,\,\alpha\ge\beta\ge\alpha/2&if&a>0,\\
\rule{0pt}{15pt}
\bullet&\beta>1/2&if&a=0,\\
\rule{0pt}{15pt}
\bullet&\alpha>1,\,\alpha\ge\beta\ge(\alpha+1)/2&if&a<0.
\end{array}
\right\}
\end{equation}

Consider the product
$$\Omega_+\,=\,\Omega_{-}\otimes C([0,\infty)),
\quad
{\cal F}_+\,=\,{\cal F}_{-}\otimes{\cal B}(C([0,\infty))),
\quad
\qq\,=\,\pp_{\hspace{-3pt}-}\otimes\pp_{\hspace{-2pt}W}$$
of a probability space 
$({\Omega}_{-},{\cal F}_{-},{\pp}_{\hspace{-3pt}-})$
and the standard Wiener space 
$(C([0,\infty))$, ${\cal B}(C([0,\infty))),\pp_{\hspace{-2pt}W})$,
and let $(x_-,y_-)$ be a given pair of random variables on
$({\Omega}_{-},{\cal F}_{-},{\pp}_{\hspace{-3pt}-})$ 
satisfying 
$\pp_{\hspace{-3pt}-}(\{x_{-}>0\})=1$.
Extend all random variables on either 
$({\Omega}_{-},{\cal F}_{-},{\pp}_{\hspace{-3pt}-})$ or
$(C([0,\infty)),{\cal B}(C([0,\infty)))$, $\pp_{\hspace{-2pt}W})$
to $({\Omega}_+,{\cal F}_+,\qq)$ 
in the canonical way without changing their notation. 

Then there exists 
a filtration $\NF_+\,=\,[{\cal F}_t^+,\,t\ge 0]$ of
sub-$\sigma$-algebras of ${\cal F}_+$,
a probability measure
$\pp_{\hspace{-3pt}+}$ on $(\Omega_+,{\cal F}_\infty^+)$,
 a pair 
$[x_+(t),y_+(t),\,t\ge 0]$
of continuous $\NF_+$-adapted processes,
and an $\NF_+$-Wiener-process $W_+$,
such that 
\begin{itemize}
\item[]
$\pp_{\hspace{-3pt}+}=\qq$ on ${\cal F}_0^+$,
\end{itemize}
and
\[
\begin{array}{rcl}
x_+(t)&>&0,\\
x_+(t)&=&\rule{0pt}{15pt}\displaystyle
x_{-}\,+\int_0^t y_+(s)\,\dd s,\\
{y_+}(t)&=&\rule{0pt}{15pt}\displaystyle
y_{-}\,+\,
a\int_0^t\frac{\dd s}{x_+(s)^\alpha}
+\sqrt{2T\gamma}\int_0^t\frac{\dd W_+(s)}{x_+(s)^\beta}
-\gamma\int_0^t\frac{y_+(s)\,\dd s}{x_+(s)^{2\beta}}\,,
\end{array}
\]
for all $t\in[0,1]$, $\pp_{\hspace{-3pt}+}$-a.s.
Note that ${\cal F}_\infty^+$ can be strictly smaller than ${\cal F}_+$.
\end{lemma}

The above result can be used to prove existence of a
global weak solution in $\NH$
which, by standard arguments, 
turns into a unique strong solution,
as the equation's coefficients 
are locally Lipschitz on $\NH$.
\begin{theorem}\label{strong+}
Assume $\gamma,T>0$, and (\ref{?}). Then,
equation (\ref{strato diff}) has a unique global strong solution in $\NH$.
\end{theorem}

In the proof of the above theorem, the global solution is constructed
by patching together solutions on finite intervals as constructed
in the proof of Lemma \ref{technical}. 
However, when $a=0$, the proof of Lemma \ref{technical} can be used
to obtain a global solution without patching, and we are going to show 
that the $x$-component of this global solution is transient.
Using a comparison argument, we can also verify this transient behaviour
in the case of $a>0$, leading to
\begin{proposition}\label{comparison}
Assume $\gamma,T>0$, and (\ref{?}). 
If $a\ge 0$, then the $x$-component of any global solution
to (\ref{strato diff}) satisfies $\lim_{t\to\infty}x(t)=+\infty$, a.s.
\end{proposition}

Of course, if a process $[{x}(t),{y}(t),\,t\ge 0]$ taking values in $\NH$,
when started at any initial condition,
almost surely satisfies $\lim_{t\to\infty}x(t)=+\infty$, 
then its dynamics would not allow for an invariant measure.
As the Lyapunov functions constructed in \cite{mattingly,funaki}
guarantee the existence of an invariant measure,
we also have the following
\begin{corollary}
Recall ${\cal L}$ given by (\ref{infi op}) with $\gamma,T>0$, 
and assume both $\beta>1/2$ as well as $\alpha\ge\beta\ge\alpha/2$.
If the potential $U$ is repulsive at zero, i.e.\ $a>0$, 
then ${\cal L}$ does not admit
Lyapunov functions of the types used in \cite{mattingly,funaki}.
\end{corollary}

However, if $a<0$ and $\alpha>1$, Theorem \ref{strong+} gives a condition
on the relationship between $\alpha$ and $\beta$ ensuring existence of a global
solution, regardless the values taken by $\gamma,T>0$.
But, we cannot tell whether this global solution $[{x}(t),{y}(t),\,t\ge 0]$
satisfies $\lim_{t\to\infty}x(t)=+\infty$,
or $\lim_{t\to\infty}x(t)=0$, 
or whether $x$ is recurrent---all three scenarios might be possible, 
depending on the choice of $\gamma,T>0$, which remains an open problem.
So far we only know from (\ref{not transient})
in the proof of Proposition \ref{comparison} that
\[
\mbox{$\lim_{t\to\infty}$}\,x(t)=+\infty,
\quad\mbox{on $\{\int_0^\infty x(t)^{-2\beta}\,\dd t\,<\,\infty\}$},
\quad\mbox{a.s.},
\]
but this set might have measure zero in cases where $a<0$.

Finally, we return to our motivating example, (\ref{single equ}), where 
$\alpha=3$ and $\beta=2$, so that $\beta=(\alpha+1)/2$,
and hence Theorem \ref{strong+} implies that (\ref{single equ})
always has a positive global solution, regardless the values
taken by $\delta\in\NR$ and $\gamma,D>0$.
\begin{corollary}
If $x$ solves (\ref{single equ}), then the non-linear wave $\psi$
given by (\ref{trial wave}) would never collapse.
\end{corollary}
\section{Discussion of Conditions}\label{why}
In this section we relate crucial steps in our proofs 
to the conditions they rely upon, which sheds some light on
how essential these conditions actually are.

First, the base step of our construction consists in analysing functionals of
the solution to equation (\ref{equ xhat}), for a given
Ornstein-Uhlenbeck process $[\hat{y}(t),\,t\ge 0]$.

If $\beta\in[0,1/2]$ then Lemma \ref{Tlimit} would be wrong 
because, almost surely, the solution to (\ref{equ xhat}) 
would be a continuous function on the compact interval $[0,\tau]$.
We do not see an easy fix for our proofs without a valid Lemma \ref{Tlimit},
and that is the main reason
why we want $\beta$ to be greater than $1/2$ throughout this paper.

Second, when $a\not=0$, the base step of our construction is followed 
by a Girsanov transform, and we have to check Novikov's condition.
Checking this condition is based on H\"older's inequality which 
can only be applied if $\alpha$ and $\beta$ 
are in the relation $\alpha\ge\beta\ge\alpha/2$.

The third and last crucial step of our construction is the proof of 
Lemma \ref{Tlimit+}. It turns out that, if $a>0$, the conditions $\beta>1/2$ and 
$\alpha\ge\beta\ge\alpha/2$ assumed in the first and second step, respectively,
are sufficient for this proof. But, if $a<0$, two extra conditions,
$\alpha>1$ and $\beta\ge(\alpha+1)/2$, are required.
Note that these extra conditions imply $\beta>1/2$ needed in the first step.

Furthermore, compared with $\alpha>1/2$, which is the consequence of condition 
$\beta>1/2,\,\alpha\ge\beta\ge\alpha/2$ from case $a>0$,
the extra condition $\alpha>1$ is in a way counter intuitive for
negative $a$. Indeed, for negative $a$, a greater power of $\alpha$
in (\ref{strato diff}) should push the trajectories of the $x$-component
further to zero, once $x$ is close to collapse. 
This push is obviously compensated by stronger fluctuations of the damped noise
caused by the other extra condition $\beta\ge(\alpha+1)/2$.
The message of the proof of Lemma \ref{Tlimit+} seems to be that
global solutions to (\ref{strato diff}) can only exist 
if $\alpha$ and $\beta$ are in the right ratio.

We finally discuss the most interesting case $\delta<0$ of our main application,
equation (\ref{single equ}), where $\alpha=3$ and $\beta=2$, so that
$\beta=(\alpha+1)/2$, which is at the `edge' of the condition
ensuring global existence. This could mean that, when $\delta<0$,
solutions to (\ref{single equ}) are `just' global in the sense that
$\lim_{t\to\infty}x(t)=0$, almost surely or with positive probability,
depending on the choice of $\gamma,D>0$.
This behaviour, which the authors called `pseudo-collapse',
has been conjectured and supported by numerical experiments
in \cite{main}.

However, when $\delta\ge 0$, we know from Proposition \ref{comparison}
that solutions are even transient and cannot pseudo-collapse,
and hence $\delta\ge 0$ would be a condition for non-pseudo-collapse
of the corresponding non-linear wave given by (\ref{trial wave}).
Recalling the definition of $\delta$ in Remark \ref{parameters}, 
this condition would read
\[
\|\psi(\cdot,0)\|_{L^2}^2
\,\le\,
2c_f^{1,0,2}\,c_f^{1,2,0}/c_f^{1,4,0},
\]
which compares the $L^2$-norm of the wave's initial condition
with a product of integrals of $f$ and $f'$.

A similar but structurally easier condition  
is well-known for the classical focusing non-linear
Schr\"odinger equation: Weinstein's criterion, \cite[Thm.\ A]{weinstein}, says that
solutions would never blow up if
\[
\|\psi(\cdot,0)\|_{L^2}\,<\,\|Q\|_{L^2},
\]
where $Q$ is the ground state used for (\ref{special blow up}).
Here, $Q$ satisfies an equation, 
while $f$ used for (\ref{trial wave}) does not.
Therefore, the condition for non-pseudo-collapse could be relaxed to
\[
\|\psi(\cdot,0)\|_{L^2}^2
\,\le\,
\sup_f\,
2c_f^{1,0,2}\,c_f^{1,2,0}/c_f^{1,4,0},
\]
where the supremum is taken over all
smooth functions $f:\NR\to(0,\infty)$ which are rapidly decreasing.
\section{Proofs}
\begin{proof}[Proof of Lemma \ref{technical}]
First, observe that $\beta>1/2$ in all three cases of (\ref{?}).

Let $[B_t,\,t\ge 0]$ be the coordinate process on 
$(C([0,\infty))$, ${\cal B}(C([0,\infty))),\pp_{\hspace{-2pt}W})$,
and define the filtration
\[
\NG_+\,=\,[{\cal G}_t^+,\,t\ge 0]
\quad\mbox{by}\quad
{\cal G}_t^+\,=\,
\sigma(x_-,y_-)
\vee
\sigma(\{B_s:s\le t\}).
\]
Note that $[B_t,\,t\ge 0]$ is a $\NG_+$-Wiener-process
on $({\Omega}_+,{\cal F}_+,\qq)$.

Of course,
$$\hat{y}(t)\,\stackrel{\mbox{\tiny def}}{=}\,e^{-\gamma t}y_- 
+
\sqrt{2T\gamma}\int_0^t e^{-\gamma(t-s)}\,\dd B_s,
\quad t\ge 0,$$
satisfies
\begin{equation}\label{equ yhat}
\hat{y}(t)\,=\,y_-
+\sqrt{2T\gamma}\,B_t-\gamma\int_0^t\hat{y}(s)\,\dd s,
\end{equation}
for all $t\ge 0$, $\qq$-a.s.

Next, for $\beta>1/2$,
\[
\hat{x}(t)\,\stackrel{\mbox{\tiny def}}{=}\,
\left[\rule{0pt}{12pt}\right.
x_-^{1-2\beta}-(2\beta-1)\int_0^t\hat{y}(s)\,\dd s
\left.\rule{0pt}{12pt}\right]^{\frac{-1}{2\beta-1}}
\]
solves
\begin{equation}\label{equ xhat}
\hat{x}(t)\,=\,x_-\,+\int_0^t\hat{x}(s)^{2\beta}\,\hat{y}(s)\,\dd s,
\quad\mbox{for $t<\tau$},
\end{equation}
where
\[
\tau\stackrel{\mbox{\tiny def}}{=}
\inf\{s\ge 0:\int_0^s\hat{y}(r)\,\dd r=\frac{x_-^{1-2\beta}}{2\beta-1}\}.
\]
\begin{lemma}\label{tauFinite}
The $\NG_+$-stopping time $\tau$ satisfies $\qq(\{\tau<\infty\})=1$.
\end{lemma}
\begin{proof} 
Rewrite (\ref{equ yhat}) to obtain
\[
\gamma\int_0^s\hat{y}(r)\,\dd r\,=\,y_- +\sqrt{2T\gamma}\,B_s-\hat{y}(s),
\quad s\ge 0,\;\mbox{$\qq$-a.s.},
\]
where
\[
\sqrt{2T\gamma}\,B_s-\hat{y}(s)
\,=\,
-e^{-\gamma s}y_-
+
\sqrt{2T\gamma}
\int_0^s[1-e^{-\gamma(s-r)}]\,\dd B_r
\]
by definition of $\hat{y}(s)$. 
Thus, since $\lim_{s\to\infty}e^{-\gamma s}y_-=0$, 
the process $s\mapsto\int_0^s\hat{y}(r)\,\dd r$ is almost surely
going to hit $x_-^{1-2\beta}/(2\beta-1)$ in finite time,
if the process $s\mapsto\int_0^s[1-e^{-\gamma(s-r)}]\,\dd B_r$ is recurrent.

To show recurrence of this process, we use the representation,
\[
\int_0^s[1-e^{-\gamma(s-r)}]\,\dd B_r
\,=\,
B_s-e^{-\gamma s}\,\tilde{B}(\frac{e^{2\gamma s}-1}{2\gamma}),
\quad s\ge 0,\;\mbox{$\qq$-a.s.},
\]
where $[\tilde{B}(t),\,t\ge 0]$ is another Wiener process 
on $({\Omega}_+,{\cal F}_+,\qq)$---see Thm.II.7.2 in \cite{IW1981}.
Taking into account the law of iterated logarithm
(cf.\ Thm.2.9.23 in \cite{KS1991}),
i.e.,
for any one-dimensional standard Wiener process $[W_t,\,t\ge 0]$,
\[
\limsup_{t\to\infty}\frac{W_t}{\sqrt{2t\log\log t}}=1,\;\mbox{a.s.},
\quad
\liminf_{t\to\infty}\frac{W_t}{\sqrt{2t\log\log t}}=-1,\;\mbox{a.s.};
\]
we can then conclude that the process
\[
s\mapsto B_s-e^{-\gamma s}\,\tilde{B}(\frac{e^{2\gamma s}-1}{2\gamma})
\]
is recurrent if
\[
\sqrt{2s\log\log s}\,
-e^{-\gamma s}\sqrt{\frac{e^{2\gamma s}-1}{\gamma}\log\log\frac{e^{2\gamma s}-1}{2\gamma}}
\]
diverges, when $s$ goes to infinity, which is true.
\end{proof}
Now introduce $[T_t,\,t\ge 0]$ defined by
$$T_t\,\stackrel{\mbox{\tiny def}}{=}\left\{\begin{array}{ccc}
\int_0^t\hat{x}(s)^{2\beta}\,\dd s&:&t<\tau,\\
\rule{0pt}{12pt}+\infty&:&t\ge\tau,
\end{array}\right.$$
which gives an increasing right-continuous $\NG_+$-adapted process.
\begin{lemma}\label{Tlimit}
For $\beta>1/2$,\hspace{3pt}
$\qq(\{\lim_{t\uparrow\tau}T_t\,=\,+\infty\})\,=\,1$.
\end{lemma}
\begin{proof}
There exists ${\Omega}_0\in{\cal F}_+$ such that
$\qq({\Omega}_0)=1$ and both $\hat{y}(s,\omega)$ is continuous in $s$
as well as $\tau(\omega)<\infty$ for all $\omega\in{\Omega}_0$. 
Choose $\omega\in{\Omega}_0$ and assume that
$$\lim_{t\uparrow\tau(\omega)}T_t(\omega)\,=\,c\,<\,+\infty.$$
Then, for $t<\tau(\omega)$, it follows from the mean value theorem that
$$c\,-T_t(\omega)\,=\,(\tau(\omega)-t)
\left[\rule{0pt}{12pt}\right.
x_-^{1-2\beta}-(2\beta-1)\int_0^{\tilde{t}}\hat{y}(s,\omega)\,\dd s
\left.\rule{0pt}{12pt}\right]^{\frac{-2\beta}{2\beta-1}}$$
where $\tilde{t}\in(t,\tau(\omega))$. Also, by definition of $\tau$ and
continuity of $\hat{y}(s,\omega)$ in $s$,
$$x_-^{1-2\beta}-(2\beta-1)\int_0^{\tilde{t}}\hat{y}(s,\omega)\,\dd s
\,=\,(2\beta-1)(\tau(\omega)-\tilde{t}\,)
\;\hat{y}\!\left(\tilde{\tilde{t}},\omega\right)$$
for some $\tilde{\tilde{t}}\in(\,\tilde{t},\tau(\omega))$, again applying
the mean value theorem. Thus,
$$c\,-T_t(\omega)\,=\,O\left((\tau(\omega)-t)^{\frac{-1}{2\beta-1}}\right),
\quad t\uparrow\tau(\omega),$$
which means that $c\,-T_t(\omega)$ should blow up when $t$ goes to $\tau(\omega)$,
since $\beta>1/2$. But, such a blow-up would contradict our assumption of $c<\infty$,
proving $\lim_{t\uparrow\tau(\omega)}T_t(\omega)=+\infty$,
for all $\omega\in{\Omega}_0$.
\end{proof}
As $[T_t,\,t\ge 0]$ is continuous and strictly increasing on $[0,\tau)$, 
Lemma \ref{Tlimit} implies that 
the right-inverse $A=[A_t,\,t\ge 0]$ defined by
$$A_t\,=\,\inf\{s\ge 0:T_s>t\}$$
is a continuous strictly increasing family of $\NG_+$-stopping times
satisfying
$$A_t\,<\,\tau,\quad t\ge 0,\,\mbox{$\qq$-a.s.}$$
As a consequence, the time-changed processes
$x_+(t)=\hat{x}(A_t),\,y_+(t)=\hat{y}(A_t)$ are
well-defined for all $t\ge 0$ 
on a set in ${\cal F}_+$ of $\qq$-measure one.
Furthermore, by time-change, equation (\ref{equ xhat}) yields
\begin{equation}\label{equ x}
x_+(t)\,=\,x_{-}\,+\int_0^t y_+(s)\,\dd s,\quad t\ge 0,\,\mbox{$\qq$-a.s.}
\end{equation}
\begin{remark}\rm\label{positive}
It follows immediately from the construction of $\hat{x}$ on
$[0,\tau)$ that
$x_+(t)>0$, for all $t\ge 0$, $\qq$-a.s.
\end{remark}
So, the first two statements of the lemma would be true under the measure $\qq$,
and, if $a=0$, by time-change, equation (\ref{equ yhat}) yields (\ref{with indicator}),
for all $t\ge 0$, \qq-a.s., in a straight forward way.
Note that $\beta>1/2$ has been the only assumption we made, so far,
proving the lemma in the case $a=0$.

In what follows, assume $a\not=0$.

Then,
the equation for $y_+(t)$ requires
a measure $\pp_{\hspace{-3pt}+}$ different to $\qq$. 
The next step is to construct this measure.

Introduce the process, $\rho=[\rho(t),\,t\ge 0]$, given by
$$\rho(t)\,\stackrel{\mbox{\tiny def}}{=}\,\exp\{
\frac{a}{\sqrt{2T\gamma}}\int_0^{t}{\bf 1}_{[0,A_1]}(s)\hat{x}(s)^{2\beta-\alpha}\,\dd B_s
-\frac{a^2}{4T\gamma}\int_0^{t}{\bf 1}_{[0,A_1]}(s)\hat{x}(s)^{4\beta-2\alpha}\,\dd s
\},$$
which is well-defined since the stochastic integrand is a
caglad $\NG_+$-\,adapted process. 
Since $\alpha\ge\beta\ge\alpha/2$, by H\"older's inequality,
\[
\int_0^{t}{\bf 1}_{[0,A_1]}(s)\hat{x}(s)^{4\beta-2\alpha}\,\dd s
\,\le\,
t^{\frac{\alpha-\beta}{\beta}}
\cdot
(\int_0^{A_1}\hat{x}(s)^{2\beta}\,\dd s)^{\frac{2\beta-\alpha}{\beta}}
\,=\,
t^{\frac{\alpha-\beta}{\beta}}\cdot (T_{A_1})^{\frac{2\beta-\alpha}{\beta}},
\]
where $\beta>1/2$ yields 
$T_{A_1}=1$, $\qq$-a.s., by Lemma \ref{Tlimit}, and hence
$$\int\exp\{\frac{a^2}{4T\gamma}
\int_0^{t}{\bf 1}_{[0,A_1]}(s)\hat{x}(s)^{4\beta-2\alpha}\,\dd s
\}\,\dd\qq\,<\,\infty,$$
for all $t\ge 0$,
so that $\rho$ is a $\NG_+$-martingale by Novikov's condition.

Copying the proof of Corollary 3.5.2 in \cite{KS1991}, one can construct
a probability measure $\pp_{\hspace{-3pt}+}$ on ${\cal G}_\infty^+$ such that
$$\hat{W}_+(t)\,\stackrel{\mbox{\tiny def}}{=}\,B_t\,-
a\int_0^t{\bf 1}_{[0,A_1]}(s)\hat{x}(s)^{2\beta-\alpha}\,\dd s/\sqrt{2T\gamma},
\quad t\ge 0,$$
is a $\NG_+$-Wiener-process.
\begin{remark}\rm\label{still ok}
\rule{1cm}{0pt}
\begin{itemize}\item[(i)]
The measure constructed in the original proof of Corollary 3.5.2 in \cite{KS1991}
would be defined on $\sigma(\{B_s:s\ge 0\})$ but our integrand 
${\bf 1}_{[0,A_1]}\hat{x}(\cdot)$ is not $\sigma(\{B_s:s\ge 0\})$-measurable.
However, the proof still works when using ${\cal G}_\infty^+$ instead.
It is not needed that $\NG_+$ satisfies the usual conditions.
\item[(ii)]
The $\sigma$-algebra ${\cal G}_\infty^+$ may be smaller than ${\cal F}_+$.
\end{itemize}
\end{remark}

The measure $\pp_{\hspace{-3pt}+}$ does not have to be absolutely continuous
w.r.t.\ $\qq$, but it is on every ${\cal G}_t^+$, where
$\pp_{\hspace{-3pt}+}
\,=\,
\rho(t)\cdot\qq$.
This allows to carry over some
$\qq$-a.s.\ events to $\pp_{\hspace{-3pt}+}$-a.s.\ events
by approximation with monotone
sequences of events, for example,
\begin{equation}\label{continuous yHat}
\pp_{\hspace{-3pt}+}
(\{\mbox{$\hat{y}(\cdot)$ continuous on $[0,\infty)$}\})\,=\,1.
\end{equation}
As a consequence,
\[
|\int_0^t\hat{y}(s)\,\dd s\;|\,<\,\infty,
\quad \mbox{for all $t\ge 0$,\quad $\pp_{\hspace{-3pt}+}$-a.s.},
\]
which yields
\[
\hat{x}(t)\,>\,0,
\quad \mbox{for all $t<\tau$,\quad $\pp_{\hspace{-3pt}+}$-a.s.}
\]
However, the results of both lemmas, \ref{tauFinite} and \ref{Tlimit}, 
might not remain true under the new measure $\pp_{\hspace{-3pt}+}$.

Indeed, though the definition of $\hat{y}$ is still the same
under $\pp_{\hspace{-3pt}+}$, the process $B$ is now a Wiener process with drift,
and hence the recurrence of the stochastic integral process used to prove Lemma \ref{tauFinite}
might fail to hold. Thus, we have to take into account a positive 
$\pp_{\hspace{-3pt}+}$-probability of the event $\{\tau=\infty\}$, 
and on this event the proof of Lemma \ref{Tlimit} does not work.
The next lemma gives conditions on the parameters $\alpha,\beta$ ensuring that
$T_\infty$ cannot be finite, on $\{\tau=\infty\}$, $\pp_{\hspace{-3pt}+}$-a.s., 
and this property turns out to be crucial for the rest of the proof.
\begin{lemma}\label{Tlimit+}
Assume $a\not=0$ and (\ref{?}). Then,
$\pp_{\hspace{-3pt}+}(\{\lim_{t\uparrow\tau}T_t=+\infty\})\,=\,1$.
\end{lemma}
\begin{proof}
First, recall (\ref{continuous yHat}) and note that, by the same principle,
(\ref{equ yhat}) is also true, for all $t\ge 0$, $\pp_{\hspace{-3pt}+}$-a.s.
Then, choose $\Omega_0$ such that $\pp_{\hspace{-3pt}+}(\Omega_0)=1$ and,
on $\Omega_0$: $\hat{y}(\cdot)$ is continuous, 
(\ref{equ yhat}) is satisfied for all $t\ge 0$,
and
\[
\limsup_{t\to\infty}\frac{\hat{W}_+(t)}{\sqrt{2t\log\log t}}=1,
\quad
\liminf_{t\to\infty}\frac{\hat{W}_+(t)}{\sqrt{2t\log\log t}}=-1.
\]
Here we used both $\beta>1/2$ and $\alpha\ge\beta\ge\alpha/2$ for validation 
of Novikov's condition to make sure that $\hat{W}_+$ as defined
above Remark \ref{still ok} is a Wiener process.

Now, choose $\omega\in\Omega_0$.
To simplify notation, for the rest of this proof, consider all random variables
being evaluated at the chosen $\omega$ without emphasising.

If $\tau<\infty$, as $\beta>1/2$,
$\lim_{t\uparrow\tau}T_t=+\infty$ can be shown
following the arguments used in the proof of Lemma \ref{Tlimit}.

If $\tau=\infty$,
using the definitions of $\hat{y}$ and $\hat{W}_+$, 
a rewrite of equation (\ref{equ yhat}) yields
\begin{align}\label{z-t}
\gamma z_t
\,=\,
y_- -e^{-\gamma t}y_-
&+
a\int_0^t[1-e^{-\gamma(t-s)}]\,\hat{x}(s)^{2\beta-\alpha}\,\dd s \nonumber \\
&+
\sqrt{2T\gamma}
\int_0^t[1-e^{-\gamma(t-s)}]\,\dd\hat{W}_+(s),
\end{align}
for all $t\ge 0$, where
\[
z_t\,
\,\stackrel{\mbox{\tiny def}}{=}\,
\int_0^t\hat{y}(s)\,\dd s,\quad t\ge 0.
\]

Since $\tau=\infty$, and since $t\mapsto z_t$ cannot explode in finite time
by our choice of $\Omega_0$, $\hat{x}(\cdot)$ is a positive function 
on the entire domain $[0,\infty)$.
Thus, simple differentiation reveals that the first integral in (\ref{z-t})
is a monotonously increasing function in $t$,
and hence the behaviour of the function $t\mapsto z_t$ will be different depending on
whether this monotone function dominates the stochastic integral or not,
when $t$ goes to infinity.

However, by our choice of $\Omega_0$, using the same arguments as in the proof of
Lemma \ref{tauFinite}, 
\[
\limsup_{t\to\infty}\frac{\int_0^t[1-e^{-\gamma(t-s)}]\,\dd\hat{W}_+(s)}{\sqrt{2t\log\log t}}=1,
\quad
\liminf_{t\to\infty}\frac{\int_0^t[1-e^{-\gamma(t-s)}]\,\dd\hat{W}_+(s)}{\sqrt{2t\log\log t}}=-1.
\]
So, if $a>0$, then the first integral in (\ref{z-t}) adds to the upward-fluctuations
of the stochastic integral leading to a finite value of $\tau$, and hence
the case $a>0$ cannot occur, once $\tau=\infty$.
Therefore, the case $a>0$ was covered above, only assuming
$\beta>1/2$ and $\alpha\ge\beta\ge\alpha/2$.

If $\tau=\infty$ and $a<0$, 
we are going to show that,
if $\alpha>1$ and $\beta>\alpha/2$, then
$T_\infty<+\infty$ would imply $\beta<(\alpha+1)/2$, 
proving the lemma in this case, too.

For the rest of this proof, 
assume $\tau=\infty,\,a<0,\,\alpha>1,\,\beta>\alpha/2,\,T_\infty<+\infty$.

To begin with, we are going to verify
that the first integral in (\ref{z-t})
would always dominate the stochastic integral, pushing all fluctuations
of $z_t$ down to $\lim_{t\to\infty}z_t=-\infty$, eventually.

Indeed, if $a<0$, when using the long-time behaviour of the
stochastic integral in (\ref{z-t}), we can deduce that, 
for some large enough $t_0$, there exists $b_0>0$ such that, for all $t\ge t_0$,
\[
\gamma z_t
\,\ge\,
y_- -e^{-\gamma t}y_-
-
|a|\int_0^t\hat{x}(s)^{2\beta-\alpha}\,\dd s 
-
b_0\sqrt{t}\sqrt{\log\log t}\,.
\]
Of course, $y_- -e^{-\gamma t}y_-$ is bigger than some negative number, for all $t\ge 0$,
and this negative number becomes even smaller when subtracting
$|a|\int_0^{t_0}\hat{x}(s)^{2\beta-\alpha}\,\dd s$.
Therefore, for all $t\ge t_0$,
the above inequality can be written as follows,
\[
z_t
\,\ge\,
-a_0-\frac{b_0}{\gamma}\sqrt{t_0}\sqrt{\log\log t_0}
-
\frac{|a|}{\gamma}\int_{t_0}^t[c_1-c_2 z_s]^{-\kappa}\,\dd s 
-
\frac{b_0}{\gamma}\int_{t_0}^t f(s)\,\dd s,
\]
writing $f(s)$ for $\frac{\dd}{\dd s}(\sqrt{t}\sqrt{\log\log t}\,)$,
and substituting the definition of $\hat{x}$, so that:
\[
c_1=x_-^{1-2\beta},\quad c_2=2\beta-1,\quad \kappa=\frac{2\beta-\alpha}{2\beta-1}.
\]

Now, consider the ordinary differential equation (ODE),
\[
\frac{\dd}{\dd t}\tilde{z}
\,=\,
-\frac{|a|}{\gamma}[c_1-c_2\tilde{z}]^{-\kappa}
-\frac{b_0}{\gamma}f(t).
\]
If this equation,
when started at $t_0$ from
$-a_0-\frac{b_0}{\gamma}\sqrt{t_0}\sqrt{\log\log t_0}$,
has a unique global solution, then,
by standard comparison arguments,
\[
z_t\,\ge\,\tilde{z}_t,\quad t\ge t_0,
\]
and thus,
\begin{equation}\label{z by p}
c_1-c_2 z_t\,\le\,p_t
\,\stackrel{\mbox{\tiny def}}{=}\,
c_1-c_2\tilde{z}_t,
\quad t\ge t_0,
\end{equation}
if $[p_t,\,t\ge t_0]$ was the unique global solution of
\begin{equation}\label{equ p}
p_t\,=\,c_1+c_2 a_0
+
c_2\frac{|a|}{\gamma}\int_{t_0}^t p_s^{-\kappa}\,\dd s
+
c_2\frac{b_0}{\gamma}\sqrt{t}\sqrt{\log\log t}\,.
\end{equation}

Yet, since $c_1+c_2 a_0>0$ and $t_0$ was chosen large enough,
this ODE (written in integral form) has local solutions,
these local solutions are unique on their domain of definition
(since the equation's coefficients are locally Lipschitz),
and any local solution is monotonously increasing.

So, on its domain of definition, any local solution satisfies
\[
p_t\ge c_2\frac{b_0}{\gamma}\sqrt{t}\sqrt{\log\log t}, 
\]
and hence, since $\kappa>0$,
\[
\int_{t_0}^t p_s^{-\kappa}\,\dd s
\,\le\,
(\frac{\gamma}{c_2 b_0})^\kappa\int_{t_0}^t s^{-\kappa/2}\,\dd s,
\]
which means that $p_t=c_1-c_2\tilde{z}_t,\,t\ge t_0$, 
is indeed the unique global solution of equation (\ref{equ p}),
because blow-up cannot occur in finite time.

Next, since $\kappa\not=2$, the above inequality asserts
\[
\int_{t_0}^t p_s^{-\kappa}\,\dd s
\,\le\,
(\frac{\gamma}{c_2 b_0})^\kappa\cdot t^{-\frac{\kappa}{2}+1},
\quad t\ge t_0,
\]
which we apply to estimate the right-hand side of (\ref{equ p}).
Here, since $\kappa<1$,
the product $\sqrt{t}\sqrt{\log\log t}$ is dominated by $t^{-\frac{\kappa}{2}+1}$,
when $t$ goes to infinity, and therefore (\ref{equ p}) yields
\[
p_t\,\le\,c_0\,t^{-\frac{\kappa}{2}+1},
\quad t\ge t_0,
\]
for some sufficiently large constant $c_0>0$.

Using this bound, (\ref{z by p}), the definition of $\hat{x}$, and $\kappa>0$,
we obtain that
\[
\frac{1}{(c_0\,t^{-\frac{\kappa}{2}+1})^\kappa}
\,\le\,
\hat{x}(t)^{2\beta-\alpha},
\quad t\ge t_0,
\]
and hence the first integral in (\ref{z-t}) is bounded below by
\[
c_0^{-\kappa}\int_{t_0}^t[1-e^{-\gamma(t-s)}]\,
s^{-\kappa(1-\kappa/2)}\,\dd s,
\quad t\ge t_0.
\]
Since $\kappa(1-\kappa/2)>0$, by l'Hospital,
\[
\lim_{t\to\infty}\int_{t_0}^t e^{-\gamma(t-s)}\,s^{-\kappa(1-\kappa/2)}\,\dd s
\,=\,0,
\]
and since $\kappa(1-\kappa/2)<1/2$,
\[
\lim_{t\to\infty}\frac{\int_{t_0}^t s^{-\kappa(1-\kappa/2)}\,\dd s}{\sqrt{2t\log\log t}}
\,=\,+\infty,
\]
finally proving our claim that the first integral in (\ref{z-t}) 
would dominate the stochastic integral, for any  $a<0$.

As a consequence, for any $a<0$, we can now conclude that
\[
\lim_{t\to\infty}z_t
\,=\,
-\lim_{t\to\infty}\int_0^t[1-e^{-\gamma(t-s)}]\,\hat{x}(s)^{2\beta-\alpha}\,\dd s
\,=\,
-\infty,
\]
and thus, by l'Hospital,
\[
\lim_{t\to\infty}\int_0^t e^{-\gamma(t-s)}\,\hat{x}(s)^{2\beta-\alpha}\,\dd s
\,=\,
\lim_{t\to\infty}\int_0^t e^{-\gamma(t-s)}\,[c_1-c_2 z_s]^{-\kappa}\,\dd s
\,=\,0.
\]
Therefore, the long-time behaviour of the right-hand side of (\ref{z-t}) 
is fully determined by the long-time behaviour of the function,
\[
t\mapsto\int_0^t\hat{x}(s)^{2\beta-\alpha}\,\dd s
\,=\,
\int_0^t[c_1-c_2 z_s]^{-\kappa}\,\dd s,
\]
and, choosing $t_0$ large enough, we can conclude that
\[
\frac{-|a|-\varepsilon}{\gamma}\int_0^t[c_1-c_2 z_s]^{-\kappa}\,\dd s
\,\le\,
z_t
\,\le\,
\frac{-|a|+\varepsilon}{\gamma}\int_0^t[c_1-c_2 z_s]^{-\kappa}\,\dd s,
\quad t\ge t_0,
\]
for some $\varepsilon>0$, such that $-|a|+\varepsilon$ is still negative,
leading to
\[
c_-\,t^{\frac{1}{\kappa+1}}
\,\le\,[c_1-c_2 z_t]\,\le\,
c_+\,t^{\frac{1}{\kappa+1}},
\quad t\ge t_0,
\]
by standard comparison arguments, where $c_+>c_->0$, of course.

Using the definition of $\hat{x}$, the above sandwich-bound translates into
\[
\frac{1}{c_+}\,t^{-\frac{2\beta}{(\kappa+1)(2\beta-1)}}
\,\le\,\hat{x}^{2\beta}(t)\,\le\,
\frac{1}{c_-}\,t^{-\frac{2\beta}{(\kappa+1)(2\beta-1)}},
\quad t\ge t_0,
\]
which means that, if $T_\infty<+\infty$, the exponent
$2\beta/(\kappa+1)/(2\beta-1)$ would have to be bigger than one, 
i.e.\ $\beta<(\alpha+1)/2$.
\end{proof}

We continue with the proof of Lemma \ref{technical}.

All in all, if $a\not=0$ and (\ref{?}),
then there exists a measure $\pp_{\hspace{-3pt}+}$ on ${\cal G}_\infty^+$
such that equation (\ref{equ x}) and Remark \ref{positive} remain true,
when the measure $\qq$ is replaced by $\pp_{\hspace{-3pt}+}$.
Furthermore, equation (\ref{equ yhat}) can be written as
\[
\hat{y}(t)\,=\,y_-\,+\,
a\int_0^t{\bf 1}_{[0,A_1]}(s)\,\hat{x}(s)^{2\beta-\alpha}\,\dd s
+\sqrt{2T\gamma}\,\hat{W}_+(t)
-\gamma\int_0^t\hat{y}(s)\,\dd s,
\]
for all $t\ge 0$, $\pp_{\hspace{-3pt}+}$-a.s., which gives
\begin{equation}\label{equ y}
{y_+}(t)\,=\,y_-\,+\,
a\int_0^t\frac{{\bf 1}_{[0,A_1]}(A_s)}{x_+(s)^\alpha}\,\dd s
+\sqrt{2T\gamma}\,\hat{W}_+(A_t)
-\gamma\int_0^t\frac{y_+(s)}{x_+(s)^{2\beta}}\,\dd s
\end{equation}
since
$$A_t\,=\,\int_0^{A_t}\hat{x}(s)^{-2\beta}\,\dd T_s
\,=\,\int_0^t x_+(s)^{-2\beta}\,\dd s,$$
for all $t\ge 0$, $\pp_{\hspace{-3pt}+}$-a.s.

Now, let ${\NF_+}$ be the time-changed filtration
given by ${\cal F}_t^+\stackrel{\mbox{\tiny def}}{=}{\cal G}_{A_t}^+,\,t\ge 0$, so that
\[
\pp_{\hspace{-3pt}+}=\qq
\quad\mbox{on}\quad
{\cal F}_0^+\,=\,{\cal G}_0^+
\]
because $\rho(0)=1$. 
Also, note that
$[x_+(t),y_+(t),\,t\ge 0]$
are ${\NF_+}$-adapted processes which are both $\pp_{\hspace{-3pt}+}$-a.s.\ continuous.
Of course, when switching to the filtration $\NF_+$, 
the measure $\pp_{\hspace{-3pt}+}$ can be restricted to ${\cal F}_\infty^+$
which might be smaller than ${\cal G}_\infty^+$.

Next, the continuous local ${\NF_+}$-martingale
$M_+(t)\stackrel{\mbox{\tiny def}}{=}\hat{W}_+(A_t),\,t\ge 0$, 
has quadratic variation $\langle M_+\rangle=A$.
Since, $\pp_{\hspace{-3pt}+}$-a.s., this quadratic variation takes the form,
$\int_0^t x_+(s)^{-2\beta}\,\dd s,\,t\ge 0$, where the integrand
$x_+(s)^{-2\beta},\,s\ge 0$, is positive and continuous, 
Theorem II.7.1 in \cite{IW1981} implies that there is an
${\NF_+}$-Wiener-process $W_+$ 
on $({\Omega}_+,{\cal F}_\infty^+,\pp_{\hspace{-3pt}+})$ 
such that
$$M_+(t)\,=\,\hat{W}_+(A_t)\,=\,
\int_0^t\frac{\dd W_+(s)}{x_+(s)^\beta},
\quad t\ge 0,\,\mbox{$\pp_{\hspace{-3pt}+}$-a.s.}$$
Hence, (\ref{equ y}) translates into
\begin{equation}\label{with indicator}
{y_+}(t)
\,=\,
y_{-}\,+\,
a\int_0^t\frac{{\bf 1}_{[0,1]}(s)}{x_+(s)^\alpha}\,\dd s
+\sqrt{2T\gamma}\int_0^t\frac{\dd W_+(s)}{x_+(s)^\beta}
-\gamma\int_0^t\frac{y_+(s)}{x_+(s)^{2\beta}}\,\dd s,
\end{equation}
for all $t\ge 0,\,\pp_{\hspace{-3pt}+}$-a.s.,
finally proving the lemma.
\end{proof}

\begin{proof}[Proof of Theorem \ref{strong+}]
As explained in Section \ref{results} in the paragraph above the theorem,
it suffices to show existence of a global weak solution.

Choose an arbitrary initial condition $(x(0),y(0))\in\NH$,
set $\Omega_-=\NR^2,\,{\cal F}_-={\cal B}(\NR^2)$,
and denote by $\pp_{\hspace{-3pt}-}$ the Dirac measure 
at the point $(x(0),y(0))$.
Let $(x_-,y_-)$ be the random variable on
$(\Omega_-,{\cal F}_-,\pp_{\hspace{-3pt}-})$ induced by the identity
on $\NR^2$.
Observe that $\pp_{\hspace{-3pt}-}(\{x_->0\})=1$ 
is an immediate consequence of $x(0)>0$.

Hence, there is a tupel
$({\Omega}_{+},{\cal F}_\infty^{+},\NF_+,{\pp}_{\hspace{-3pt}+},
[x_{+}(t)$, $y_{+}(t),W_+(t),\,t\ge 0])$
the components of which satisfy the properties stated in 
the conclusion of Lemma \ref{technical}.
Moreover, using $\dd x_+(t)=y_+(t)\,\dd t$ and (\ref{with indicator})
when multiplying $ x_+(t)^\beta$ by $y_+(t)$,
we obtain that
\begin{equation}\label{better wipro}
\begin{array}{rcl}
\sqrt{2T\gamma}\,W_+(t)
&=&\displaystyle
x_+(t)^\beta y_+(t)-x_-^\beta y_-
-\beta\int_0^t x_+(s)^{\beta-1}\,y_+(s)^2\,\dd s\\
&-&\rule{0pt}{20pt}\displaystyle
a\int_0^t\frac{{\bf 1}_{[0,1]}\,\dd s}{x_+(s)^{\alpha-\beta}}
+\gamma\int_0^t\frac{y_+(s)\,\dd s}{x_+(s)^\beta}\,,
\end{array}
\end{equation}
for all $t\ge 0$, $\pp_{\hspace{-3pt}+}$-a.s.,
and hence $W_+$ can be considered a Wiener process with respect to the filtration
$[\sigma(\{(x_+(s),y_+(s)):s\le t\}),\,t\ge 0]$.
Note that the proof of Lemma \ref{technical} makes clear that no extra sets
of measure zero have to be added to this filtration.

The next step is to construct, by induction,
a sequence 
$({\Omega}_{n},{\cal F}_{n},\NF_n,{\pp}_{\hspace{-3pt}n},
[x_{n}(t)$, $y_{n}(t),W_n(t),\,t\ge 0])$, $n=1,2,\dots$, such that
\[
\begin{array}{rcl}
x_n(t)&>&0,\\
x_n(t)&=&\rule{0pt}{20pt}\displaystyle
x(0)\,+\int_0^t y_n(s)\,\dd s,\\
{y_n}(t)&=&\rule{0pt}{20pt}\displaystyle
y(0)
+
a\int_0^t\frac{\dd s}{x_n(s)^\alpha}
+\sqrt{2T\gamma}\int_0^t\frac{\dd W_n(s)}{x_n(s)^\beta}
-\gamma\int_0^t\frac{y_n(s)\,\dd s}{x_n(s)^{2\beta}}\,,
\end{array}
\]
for all $t\in[0,n]$, $\ppn$-a.s.,
where $[x_{n}(t),y_{n}(t),\,t\ge 0]$
are continuous processes,
$\NF_n$ stands for the filtration 
${\cal F}_t^n=\sigma(\{(x_n(s),y_n(s)):s\le t\}),\,t\ge 0$,
and $[W_n(t),\,t\ge 0]$ is an $\NF_n$-Wiener process.

Observe that the tupel
$({\Omega}_{+},{\cal F}_\infty^{+},\NF_1,{\pp}_{\hspace{-3pt}+},
[x_{+}(t)$, $y_{+}(t),W_+(t),\,t\ge 0])$
found in the first part of the proof
plays the role of the initial case $n=1$, of course.

So, fix $n\ge 2$, and suppose that 
$({\Omega}_{n-1},{\cal F}_{n-1},\NF_{n-1},{\pp}_{\hspace{-3pt}n-1},
[x_{n-1}(t)$, $y_{n-1}(t)$, $W_{n-1}(t),\,t\ge 0])$ has already been constructed.

Reset $\Omega_-=\Omega_{n-1},\,{\cal F}_-={\cal F}_{n-1},\,
{\pp}_{\hspace{-3pt}-}={\pp}_{\hspace{-3pt}n-1}$,
and choose $x_-=x_{n-1}(n-1),\,y_-=y_{n-1}(n-1)$.
Then, again by Lemma \ref{technical},
there is a corresponding tupel
$({\Omega}_{+},{\cal F}_\infty^{+},\NF_+,{\pp}_{\hspace{-3pt}+}$,
$[x_{+}(t)$, $y_{+}(t),W_+(t),\,t\ge 0])$
which we now denote by
$({\Omega}_{n},{\cal F}_{n},{\NF}_{+},{\pp}_{\hspace{-3pt}n},
[{x}_{+}(t)$, ${y}_{+}(t),{W}_{+}(t),\,t\ge 0])$.
\begin{remark}\rm\label{change G}
In Lemma \ref{technical}, the filtration ${\NF}_{+}$ was given by
\[
{\cal F}_t^+\,=\,{\cal G}_{A_t}^+,
\quad\mbox{using}\quad
{\cal G}_t^+\,=\,
\sigma(x_-,y_-)
\vee
\sigma(\{B_s:s\le t\}),
\]
for the purpose of Remark \ref{still ok}.
But, when applying Lemma \ref{technical} in the context of the present proof,
we are going to work with
\[
{\cal G}_t^+\,=\,
\sigma(\{(x_{n-1}(s),y_{n-1}(s)):s\le n-1\})
\vee
\sigma(\{B_s:s\le t\})
\]
instead, without violating the truth of Remark \ref{still ok}.
\end{remark}

Recall that $[x_{n-1}(t),y_{n-1}(t)$, $W_{n-1}(t),\,t\ge 0]$ are extended to 
$(\Omega_{n},{\cal F}_{n},{\pp}_{\hspace{-3pt}n})$ 
in the canonical way without changing their notation.
Define, for $t\ge 0$,
\begin{eqnarray*}
x_{n}(t)&=&
x_{n-1}(t\wedge(n-1))+{x}_{+}((t-n+1)\vee 0)-x_{n-1}(n-1),\\
y_{n}(t)&=&
y_{n-1}(t\wedge(n-1))+{y}_{+}((t-n+1)\vee 0)-y_{n-1}(n-1),\\
W_{n}(t)&=&
W_{n-1}(t\wedge(n-1))+{W}_{+}((t-n+1)\vee 0),
\end{eqnarray*}
and build a filtration $\hat{\NF}_{n}$ from both
$\NF_{n-1}$ and ${\NF}_{+}$ by
$$\hat{\cal F}_t^{n}\,\stackrel{\mbox{\tiny def}}{=}\left\{\begin{array}{ccl}
{\cal F}_t^{n-1}&:&t<n-1,\\
\rule{0pt}{15pt}
{\cal F}_{t-n+1}^{+}&:&t\ge n-1.
\end{array}\right.$$

All in all,
because of
\[
\ppn
\,=\,
{\pp}_{\hspace{-3pt}+}\,=\,\qq
\,=\,
{\pp}_{\hspace{-3pt}-}\otimes\pp_{\hspace{-2pt}W}
\,=\,
{\pp}_{\hspace{-3pt}n-1}\otimes\pp_{\hspace{-2pt}W}
\quad\mbox{on}\quad
{\cal F}_0^+,
\]
and because ${\cal F}_t^{n-1}\subseteq{\cal F}_0^+$
(see Remark \ref{change G}), 
the processes
$[x_{n}(t),y_{n}(t),W_n(t)$, $t\ge 0]$ 
would have all properties needed for the induction step,
except that $W_n$ is a Wiener process with respect to the
filtration $\hat{\NF}_n$ which is possibly bigger than $\NF_n$.

However, as a consequence of (\ref{with indicator}), we also have
\[
{y_n}(t)
\,=\,
y(0)
+
a\int_0^t\frac{{\bf 1}_{[0,n]}\,\dd s}{x_n(s)^\alpha}
+\sqrt{2T\gamma}\int_0^t\frac{\dd W_n(s)}{x_n(s)^\beta}
-\gamma\int_0^t\frac{y_n(s)\,\dd s}{x_n(s)^{2\beta}},
\]
for all $t\ge 0$, $\ppn$-a.s., 
leading to the $n$th-step analogue of (\ref{better wipro}), i.e.,
\begin{align}\label{better wipro for n}
\sqrt{2T\gamma}\,W_n(t)
&=
x_n(t)^\beta y_n(t)-x(0)^\beta y(0)
-\beta\int_0^t x_n(s)^{\beta-1}\,y_n(s)^2\,\dd s \nonumber \\
&-\rule{0pt}{20pt}
a\int_0^t\frac{{\bf 1}_{[0,n]}\,\dd s}{x_n(s)^{\alpha-\beta}}
+\gamma\int_0^t\frac{y_n(s)\,\dd s}{x_n(s)^\beta}\,,
\end{align}
for all $t\ge 0$, $\pp_{\hspace{-3pt}n}$-a.s.,
so that $W_n$ is $\NF_n$\,-adapted,
and therefore it must be an $\NF_n$\,-Wiener process, too.

The next step of the proof consists in constructing a measure
on $((\NR^2)^{[0,\infty)}$, ${\cal B}((\NR^2)^{[0,\infty)}))$
whose finite-dimensional distributions are induced
by the laws of the two-dimensional processes 
$[x_n(t),\,y_n(t),\,t\ge 0],\,n=1,2,\dots$, in the following way.

For an arbitrary finite sequence of non-negative mutually different numbers 
${\bf t}=(t_1,\dots,t_k)$, define
$$\pp_{\bf t}(\Gamma)\,=\,
\ppn(\,
\overbrace{\{
(x_n(t_1),y_n(t_1),\dots,x_n(t_k),y_n(t_k))\in\Gamma\}
}^{cylinder\;set}
\,),
\quad\Gamma\in{\cal B}(\NR^{2k}),$$
where $n$ is the smallest integer such that $n\ge\max\{t_1,\dots,t_k\}$.
Then, $\{\pp_{\bf t}\}$ is a consistent family of finite-dimensional
distributions in the sense of Kolmogorov.
Hence, there is a probability measure $\pp$ on
$((\NR^2)^{[0,\infty)},{\cal B}((\NR^2)^{[0,\infty)}))$ satisfying
$$\pp_{\bf t}(\Gamma)\,=\,\pp(\{
(x(t_1),y(t_1),\dots,x(t_k),y(t_k))\in\Gamma\}),
\quad\Gamma\in{\cal B}(\NR^{2k}),$$
where $[x(t),\,y(t),\,t\ge 0]$ denotes
the coordinate process on $\Omega=(\NR^2)^{[0,\infty)}$.
Let ${\cal F}$ be the completion of 
${\cal B}((\NR^2)^{[0,\infty)})$ with respect to $\pp$,
and let $\NF$ be the filtration 
obtained by $\pp$-augmentation of
$\sigma(\{x(s),\,y(s):s\le t\}),\,t\ge 0$.

Note that, almost surely, $[x(t),\,y(t),\,t\ge 0]$ 
is a pair of continuous $\NF$-adapted processes.
This is seen by two simple arguments. 
First, since\footnote{As this probability can be approximated
by probabilities of cylinder sets.}
\[
\ppn(\{\mbox{$t\mapsto(x_n(t),y_n(t))$ continuous on $[0,n]$}\})
\,=\,
1,
\]
each of the events 
$\{\mbox{$t\mapsto(x(t),y(t))$ continuous on $[0,n]$}\}$
is in ${\cal F}$, $n=1,2,\dots$, and second,
\begin{align*}
&\;\pp(\{\mbox{$t\mapsto(x(t),y(t))$ continuous on $[0,n]$}\})\\
=\;&\;
\ppn(\{\mbox{$t\mapsto(x_n(t),y_n(t))$ continuous on $[0,n]$}\}),
\end{align*}
for each $n\ge 1$. 

In what follows, $[x(t),\,y(t),\,t\ge 0]$ always
stands for a fixed continuous version indistinguishable of the coordinate process.

Of course, in a similar way, one shows that 
$x(t)>0$, $t\ge 0$, $\pp$-a.s.,
as well as
\begin{equation}\label{x semi}
x(t)\,=\,x(0)\,+\int_0^t y(s)\,\dd s,
\quad t\ge 0,\,\mbox{$\pp$-a.s.}
\end{equation}

Now, introduce
\[
\begin{array}{rcl}
W(t)
&\stackrel{\mbox{\tiny def}}{=}&\displaystyle
\frac{x(t)^\beta y(t)-x(0)^\beta y(0)}{\sqrt{2T\gamma}}\,
-\beta\int_0^t\frac{x(s)^{\beta-1}\,y(s)^2}{\sqrt{2T\gamma}}\,\dd s\\
&-&\rule{0pt}{20pt}\displaystyle
a\int_0^t\frac{\dd s}{\sqrt{2T\gamma}\,x(s)^{\alpha-\beta}}
+\gamma\int_0^t\frac{y(s)\,\dd s}{\sqrt{2T\gamma}\,x(s)^\beta}\,,
\quad t\ge 0,
\end{array}
\]
and observe that this process is a Wiener process
because, by (\ref{better wipro for n}), it satisfies
$$\pp(\{(W(t_1),\dots,W(t_k))\in\Gamma\})\,=\,
\ppn(\{(W_n(t_1),\dots,W_n(t_k))\in\Gamma\}),$$
for every ${\bf t}=(t_1,\dots,t_k),\,n\ge\max\{t_1,\dots,t_k\}$
and $\Gamma\in{\cal B}(\NR^k)$,
and because the continuity of $[x(t),\,y(t),\,t\ge 0]$ makes
it a continuous process. Furthermore, as
$W_n$ can be considered a Wiener process with respect to the filtration
obtained by $\ppn$-augmentation of
$\sigma(\{(x_n(s),y_n(s)):s\le t\}),\,t\ge 0$,
one can also consider $W$ to be an $\NF$-Wiener-process.

Finally, again using the corresponding property on $(\Omega_n,{\cal F}_n,\ppn)$,
each of the processes $[y(t\wedge n),\,t\ge 0]$
is an $\NF$-semimartingale, $n=1,2,\dots$, and hence
$[y(t),\,t\ge 0]$ is one, too.
As a consequence, by partial integration,
$$\sqrt{2T\gamma}\,W(t)\,=\,
\int_0^t x(s)^\beta\,\dd y(s)
-a\int_0^t\frac{\dd s}{x(s)^{\alpha-\beta}}
+\gamma\int_0^t\frac{y(s)\,\dd s}{x(s)^\beta},
\quad t\ge 0,\,\mbox{$\pp$-a.s.},$$
follows from (\ref{x semi}) and the definition of $W(t)$. 
Thus, using the above right-hand side when calculating
$\int_0^t x(s)^{-\beta}\,\dd W(s)$ for any $t\ge 0$,
eventually proves the theorem.
\end{proof}

\begin{proof}[Proof of Proposition \ref{comparison}]
To begin with, assume $a\in\NR$, and let $[x(t), y(t),\,t\ge 0]$ be the solution
of (\ref{strato diff}), started at $(x(0),y(0))\in\NH$,
and driven by a Wiener process $[W(t),\,t\ge 0]$,
given on a probability space $(\Omega,{\cal F},\pp)$, i.e.,
\begin{align*}
x(t)\,
&=\,
x(0)+\int_0^t y(s)\,\dd s,\\
y(t)\,
&=\,
y(0)+a\int_0^t\frac{\dd s}{x(s)^\alpha}
+\sqrt{2T\gamma}\int_0^t\frac{\dd W(s)}{x(s)^\beta}
-\gamma\int_0^t\frac{y(s)}{x(s)^{2\beta}}\,\dd s,
\end{align*}
for all $t\ge 0$, \pp-a.s.

Since the stochastic integral in the above equation is well-defined
for all $t\ge 0$, its quadratic variation,
$\tilde{A}_t\,=\int_0^t x(s)^{-2\beta}\,\dd s$,
is well-defined for all $t\ge 0$, too, and 
\[
\tilde{T}_t\,=\,\inf\{s\ge 0:\tilde{A}_s>t\},\quad t\ge 0,
\]
can be used as a time-change.
Furthermore, since $x(t)>0$, for all $t\ge 0$, \pp-a.s., this time-change
is strictly increasing and continuous on $[0,\tilde{A}_\infty)$, \pp-a.s., and
\begin{equation}\label{no jump}
\pp(\{\mbox{$\lim_{t\uparrow\tilde{A}_\infty}$}\tilde{T}_t=+\infty\})\,=\,1.
\end{equation}

So, on the one hand, the time-changed processes,
$t\mapsto\tilde{x}(t)=x(\tilde{T}_t)$
and $t\mapsto\tilde{y}(t)=y(\tilde{T}_t)$,
are almost surely continuous processes on $[0,\tilde{A}_\infty)$.
On the other hand, since
\[
\tilde{T}_t\,=\,\int_0^t\tilde{x}(s)^{2\beta}\,\dd s,
\quad t<\tilde{A}_\infty,\;\mbox{\pp-a.s.},
\]
property (\ref{no jump}) implies that
\begin{equation}\label{limsup}
\mbox{$\limsup_{t\uparrow\tilde{A}_\infty}$}\,\tilde{x}(t)\,=\,+\infty,
\quad\mbox{on $\{\tilde{A}_\infty<\infty\}$},
\quad\mbox{\pp-a.s.},
\end{equation}
and we are going to show next that,
on $\{\tilde{A}_\infty<\infty\}$,
$\liminf_{t\uparrow\tilde{A}_\infty}\,\tilde{x}(t)$
cannot be finite with positive probability, either,
even if the parameter $a$ is negative.

Applying the time-change to the above equation yields
\begin{equation}\label{comp1}
\begin{array}{rcl}
\tilde{x}(t)
&=&\displaystyle
x(0)+\int_0^t\tilde{x}(s)^{2\beta}\,\tilde{y}(s)\,\dd s,\\
\tilde{y}(t)
&=&\displaystyle
y(0)+a\int_0^t\tilde{x}(s)^{2\beta-\alpha}\,\dd s
+\sqrt{2T\gamma}\,\tilde{W}(t)
-\gamma\int_0^t\tilde{y}(s)\,\dd s,
\end{array}
\end{equation}
for all $t<\tilde{A}_\infty$, \pp-a.s.,
where $[\tilde{W}(t),\,t\ge 0]$ is another Wiener process
on a possibly enlarged\footnote{By standard convention, the enlarged space
is denoted by $(\Omega,{\cal F},\pp)$, too.} probability space 
(cf.\ Theorem II.7.2' in \cite{IW1981}).

Of course, by the continuity properties of the time-changed processes,
\[
\tilde{x}(t)\,=\,
\left[\rule{0pt}{12pt}\right.
x(0)^{1-2\beta}-(2\beta-1)\int_0^t\tilde{y}(s)\,\dd s
\left.\rule{0pt}{12pt}\right]^{\frac{-1}{2\beta-1}},
\]
for at least all $t<\tilde{A}_\infty$, \pp-a.s., where
\[
\tilde{y}(s)
\,=\,
e^{-\gamma s}y(0) 
+
\sqrt{2T\gamma}\int_0^s e^{-\gamma(s-r)}\,\dd\tilde{W}(r)
+
a\int_0^s e^{-\gamma(s-r)}\,\tilde{x}(r)^{2\beta-\alpha}\,\dd r,
\]
for all $s<\tilde{A}_\infty$, \pp-a.s.

The question is now whether different sequences of time points, $(t_n)_{n=1}^\infty$,
converging to $\tilde{A}_\infty<\infty$, 
can lead to different limits of $\tilde{x}(t_n),\,n\to\infty$,
which can only happen if $\int_0^{t_n}\tilde{y}(s)\,\dd s$
has different limits for different sequences of time points,
which can only happen if $\tilde{y}(t_n)$
has different limits for different sequences of time points.

The key to the answer of this question is writing
\begin{equation}\label{tildeY}
\tilde{y}(s)
\quad\mbox{as}\quad
\hat{y}(s)+a e^{-\gamma s} R_s,
\end{equation} 
where $[\hat{y}(s),\,s<\tilde{A}_\infty]$ is indistinguishable 
of an Ornstein-Uhlenbeck process 
restricted to $[0,\tilde{A}_\infty)$, and
\[
R_s\,=\,
\int_0^s e^{\gamma r}\,\tilde{x}(r)^{2\beta-\alpha}\,\dd r,
\quad s<\tilde{A}_\infty.
\]
Note that $s\mapsto R_s$ is almost surely a continuous monotone function
on $[0,\tilde{A}_\infty)$, so that $a e^{-\gamma t_n} R_{t_n}$
can only have one limit for any sequence of time points
converging to $\tilde{A}_\infty$, on $\{\tilde{A}_\infty<\infty\}$, \pp-a.s.,
and the same applies to $\hat{y}(t_n)$, as this function
can almost surely be extended to a continuous function on $[0,\infty)$.

All in all, 
$\liminf_{t\uparrow\tilde{A}_\infty}\,\tilde{x}(t)$
has indeed to coincide with
$\limsup_{t\uparrow\tilde{A}_\infty}\,\tilde{x}(t)$,
on $\{\tilde{A}_\infty<\infty\}$, \pp-a.s., proving
\begin{equation}\label{not transient}
\mbox{$\lim_{t\to\infty}$}\,{x}(t)\,=\,+\infty,
\quad\mbox{on $\{\tilde{A}_\infty<\infty\}$},
\quad\mbox{\pp-a.s.},
\end{equation}
by (\ref{limsup}), for any parameter $a\in\NR$,
because $x(t)=\tilde{x}(\tilde{A}_t),\,t\ge 0$, \pp-a.s.

In the second part of the proof, we will show that the event
$\{\tilde{A}_\infty<\infty\}$ has probability one,
for any $a\ge 0$, eventually proving the proposition.

First, the Ornstein-Uhlenbeck process used in (\ref{tildeY})
and the process $[\hat{y}(t),\,t\ge 0]$ used in the proof of Lemma \ref{technical}
have the same law, when started at $y(0)$, justifying the same notation.
Therefore,
\begin{align*}
\hat{x}(t)\,
&=
\left[\rule{0pt}{12pt}\right.
x(0)^{1-2\beta}-(2\beta-1)\int_0^t\hat{y}(s)\,\dd s
\left.\rule{0pt}{12pt}\right]^{\frac{-1}{2\beta-1}},\;t<\tau,\\
\tau\,
&=\,
\inf\{s\ge 0:\int_0^s\hat{y}(r)\,\dd r=\frac{x(0)^{1-2\beta}}{2\beta-1}\},
\end{align*}
and the corresponding objects given in the proof of Lemma \ref{technical}
have the same law, too, when using $x(0)$ instead of $x_-$, so that
\begin{equation}\label{tau finite again}
\pp(\{\tau<\infty\}
\cap
\{\mbox{$\lim_{t\uparrow\tau}$}\,\hat{x}(t)\,=\,+\infty\})
\,=\,1
\end{equation}
can easily be derived from Lemma \ref{tauFinite}. Furthermore,
\begin{equation}\label{comp2}
\begin{array}{rcl}
\hat{x}(t)
&=&\displaystyle
x(0)+\int_0^t\hat{x}(s)^{2\beta}\,\hat{y}(s)\,\dd s,\\
\hat{y}(t)
&=&\displaystyle
y(0)+\sqrt{2T\gamma}\,\tilde{W}(t)
-\gamma\int_0^t\hat{y}(s)\,\dd s,
\end{array}
\end{equation}
for all $t<\tau$, \pp-a.s.

The next step is to show that, if $a\ge 0$, then
$(\tilde{x}(t),\tilde{y}(t))$ satisfying (\ref{comp1})
dominates
$(\hat{x}(t),\hat{y}(t))$ satisfying (\ref{comp2}),
for all $t<\tau\wedge\tilde{A}_\infty$.
In fact, since the drift coefficient of equation (\ref{comp2}),
i.e.\ $(\hat{x},\hat{y})\mapsto(\hat{x}^{2\beta}\,\hat{y},-\gamma\hat{y})$,
is quasi-monotonously increasing (cf.\ Def.3.1 in \cite{assing}),
and since solutions to (\ref{comp2}) are pathwise unique
up to $\tau\wedge\tilde{A}_\infty$,
and since the difference of the  drift coefficients of (\ref{comp1}) and (\ref{comp2})
is a vector field on $\NH$ with non-negative components,
it follows from Prop.3.3 in \cite{assing} that
\[
\tilde{x}(t)\ge\hat{x}(t),\quad 
\tilde{y}(t)\ge\hat{y}(t),\quad
t<\tau\wedge\tilde{A}_\infty,\,\mbox{\pp-a.s.}
\]
\begin{remark}\rm
The results in \cite{assing} were obtained for coefficients defined on $\NR^d$,
but it is an easy exercise to show their validity for coefficients defined
on domains like our half-plane, $\NH$.
\end{remark}

Now, if $\tilde{A}_\infty$ was bigger than $\tau$ with positive probability,
then the process $t\mapsto\tilde{x}(t)$, being continuous on $[0,\tilde{A}_\infty)$,
would blow up before $\tilde{A}_\infty$, by (\ref{tau finite again}), 
which is a contradiction. Thus, $\tilde{A}_\infty$ is almost surely bounded
by $\tau$ from above leading to $\pp(\{\tilde{A}_\infty<\infty\})=1$,
again by (\ref{tau finite again}).
\end{proof}
\newpage
%
%

\end{document}